\documentclass[11pt, letterpaper, onecolumn, oneside, final]{article}

\usepackage[left=1in,right=1in,top=1in,bottom=1in]{geometry}

\usepackage{xcolor}
\usepackage{graphicx}

\usepackage{amssymb}
\usepackage{amsmath}

\usepackage{amsthm}
\theoremstyle{plain}

\newtheorem{theorem}{Theorem}

\newtheorem{lemma}{Lemma}
\theoremstyle{definition}

\newtheorem{definition}{Definition}

\newtheorem{assumption}{Assumption}

\newcommand{\set}[1]{\left\{#1\right\}}

\newcommand{\pf}{\noindent \textbf{Proof}: }
\newcommand{\pfend}{\hfill $\blacksquare$}

\begin{document}

\title{\bf Inversion-based Measurement of Data Consistency for Read/Write Registers
}



\author{
    Yu Huang, Hengfeng Wei, Maosen Huang, Lingzhi Ouyang\\
    {\small State Key Laboratory for Novel Software Technology, Nanjing University} \\
    {\small Department of Computer Science and Technology, Nanjing University} \\
    {\footnotesize\textsf{\{yuhuang, hfwei\}@nju.edu.cn, methor1992@gmail.com, lingzhi.ouyang@outlook.com}}
}

\date{}

\maketitle

\begin{abstract}

\normalsize

Both providers and consumers of distributed storage services benefit from the quantification of the severity of consistency violations. However, existing methods fail to capture a typical pattern of violation - the disorder among operations with reference to the ideally strong sequential execution. Such disorder is often seen in Internet applications based on distributed storage services, such as instant messaging. To this end, we use inversions in a permutation with reference to the ideally strong sequential execution to measure the consistency of data. The $i$-atomicity model is defined and a (pseudo-)polynomial verification algorithm for a restricted type of histories is proposed. The basic idea of the verification algorithm is brute-force search for the permutation with less than $i$ inversions, which requires exponential cost. Reasonable assumptions abstracted from application scenarios can be leveraged to prune the search space, which yields an efficient polynomial verification algorithm.
   
\end{abstract}


\section{Introduction}

Distributed storage systems have become commercially popular and enable a variety of Internet services today \cite{Chang08, DeCandia07, Cooper08, Beaver10}. Such systems are expected to be fast, always available, highly scalable, and partition-tolerant. To this end, they typically offer simple key-value operations, and replicate data across machines or even data-centers, at the expense of introducing data inconsistency \cite{Brewer00, Bermbach14}.

Consistency is a system quality which requires a level of concern comparable to performance and availability, e.g., in the CAP principle \cite{Brewer00} and in the BASE semantics \cite{Fox97}. Though practitioners widely use the notion of \textit{eventual consistency}, this is a rather fuzzy consistency term and covers a broad range of actual consistency behaviors \cite{Bermbach14}.  Both providers and consumers of the distributed storage service benefit from knowing the precise degree of data (in)consistency.

On the one hand, the service providers need to know what degree of consistency is actually provided to the consumers and how much burden it requires to develop upper layer applications. Then they can find room for improvements, possibly without compromising performance or availability. On the other hand, when the service consumers know better the degree of data inconsistency, the handling of data update conflicts may be significantly simplified. Measurement of data consistency enables both parties to better negotiate the service level agreement, and monetary compensations proportional to the measurement can be supported \cite{Golab11}.

Existing techniques for data consistency measurement, mainly time-based \cite{Golab11, Golab14} and version-based \cite{Golab13, Golab18}, fail to capture one type of anomaly often seen in Internet services, such as instant messaging \cite{Zheng17}. In instant messaging scenarios, one salient type of anomaly is the disorder among messages. While most messages can be delivered in time, often some messages may be out of order and possibly are reordered after delivery, especially in multi-party and multi-device scenarios \footnote{In multi-party scenarios, the conversation usually involves at least three people, while in multi-device scenarios, the same account may log in on multiple devices e.g., PCs, laptops, pads and mobile phones.}.

To cope with this problem, we employ the notion of \textit{inversion} to quantify the severity of consistency violation. Specifically, an inversion indicates the disorder between two elements with reference to a given total oder \cite{Knuth98}. It is used to define the consistency model we call $i$-\textit{atomicity}. When an execution is said to be $i$-atomic, it principally means that operations in the execution can be arranged in a permutation, and no more than $i$ inversions are incurred with reference to the total order required by atomicity. Here, atomicity is an ideally strong consistency condition, requiring the execution to be equivalent to a legal sequential execution \cite{Lamport86a, Herlihy90}. In the $i$-atomicity model, the severity of consistency violation is defined as the extent to which operations are out-of-order. The disorder is measured by the number of inversions incurred.

The definition $i$-atomicity naturally raises the $i$-AV problem, i.e. verifying whether a given history of executions (log of shared register accesses) satisfies $i$-atomicity. The $i$-AV problem is NP-Complete in the general case assuming that $i$ is part of the input, since it generalizes the atomicity verification problem \cite{Gibbons97}. We propose a pseudo-polynomial algorithm for a restricted type of histories. Our verification algorithm enumerates all possible permutations based on a structure we call the configuration graph (CG). The $i$-AV problem is solved by the traversal of CG at exponential cost. Reasonable assumptions abstracted from practical scenarios can be leveraged to prune the CG, leading to efficient traversal in polynomial time.

The rest of this work is organized as follows. Section \ref{Sec:Related-Work} overviews the existing work. Section \ref{Sec:Preli} and \ref{Sec:Alg} present the $i$-atomicity model and the verification algorithm. Finally, Section \ref{Sec:Conclu} concludes the paper with a summary and discussions on the future work.

\section{Related work} \label{Sec:Related-Work}

Distributed storage systems emphasize the paramount importance of high availability and resilience to failures \cite{DeCandia07, Cooper08}. Thus eventual consistency is often adopted \cite{Bermbach14}. The notion of eventual consistency is promoted by practitioners. It is different from the rigorous and formal treatment of consistency models in the academia \cite{Steinke04}.

The fuzzy nature of eventual consistency and its wide adoption motivate the quantification of severity of consistency violations \cite{Golab11, Golab13, Golab14, Golab18}. Existing quantification techniques can be viewed from two perspectives. From the micro-perspective, time-based quantification techniques, e.g. $\Delta$-atomicity \cite{Golab11} and $\Gamma$-atomicity \cite{Golab14}, are adopted for applications sensitive to realtime consistency requirements. Time-based techniques measure data consistency by how many time units the value returned has been stale. For applications sensitive to the frequency of data updates, version-based quantification techniques , e.g. $k$-atomicity \cite{Aiyer05, Golab18}, are proposed. Version-based techniques measure data consistency by how many times the data has been updated when the returned value is obtained. From the macro-perspective, Yu and Vahdat propose TACT (Tunable Availability and Consistency Tradeoffs), a middleware layer that supports application-dependent consistency semantics expressed as a vector of metrics defined over a logical consistency unit or conit \cite{Yu02}. Inconsistencies in the observed value of a conit are bounded using three metrics: numerical error, order error, and staleness. 


The notion of $i$-atomicity proposed in this work is inspired by the $k$-atomicity model and the order error in the TACT framework. However, existing metrics for consistency measurement cannot delineate the disorder among operations. The order error in TACT is defined as the weighted out-of-order writes that affect a conit. The $i$-atomicity model puts emphasis on the permutation of all operations as a whole, which naturally extends the definition of atomicity.

Both $k$-AV and $i$-AV are NP-Complete in the general case, since they include the verification of atomicity as its special case. Read-mapping is used to circumvent the NP-Completeness of atomicity verification \cite{Gibbons97}. Assuming read-mapping, the $k=2$ case for $k$-AV is solved in \cite{Golab14}. For $k\geq 3$, the $k$-AV problem is solved only for special cases. Similarly, we also assume the existence of read-mapping, and solve the $i$-AV problem for a restricted type of histories. For fixed $k$ and $i$, the problem of proving NP-Completeness for $k$-AV and $i$-AV for are principally open.

\section{Inversion-based Measurement of Data Consistency} \label{Sec:Preli}

\subsection{Preliminaries}

We model a distributed key-value store as a collection of read/write registers, replicated over multiple servers, as in \cite{Golab18}. We define an execution history (or history for short) as a sequence of events where each event is either the invocation or the response of an operation on a read/write register. Each operation in the history has both an invocation and a response event. The total number of operations and the number of writes in the input history are denoted by $n$ and $n_w$, respectively. As for shared objects that appear in the history, we assume that:
\begin{assumption}[Single shared object] \label{Asmp:Single}
    All operations in the history are applied to the same object.
\end{assumption}

\noindent Note that Assumption \ref{Asmp:Single} is restrictive. See further discussions on the locality of $i$-atomicity in Section \ref{Sec:Conclu}.

Each event in the history is tagged with a unique time, and events appear in the history in increasing order of their timestamps. 
Given a read $r$ and a write $w$, we call $w$ a dictating write of $r$ if $r$ and $w$ share the same value. In that case we call $r$ a dictated read of $w$. A write may have any number of dictated reads, but for the number of dictating write we assume that:
\begin{assumption}[Read-mapping]
    Every read operation is mapped to its unique dictating write.
\end{assumption}

For history $\sigma$, we can define the partial order between operations. Let $o.s$ and $o.f$ denote the timestamps of the invocation and the response events of operation $o$ respectively, and define $o_1 \rightarrow o_2$ if $o_1.f < o_2.s$. We define $o_1 || o_2$ if neither $o_1 \rightarrow o_2$ nor $o_2 \rightarrow o_1$ holds. For the ease of presentation, we assume that:
\begin{assumption}[Read after write] \label{Asmp:Read-After}
    If $r$ is a read operation and $w$ is its dictating write, then $r \not \rightarrow w$ (i.e., $w\rightarrow r$ or $w||r$).
\end{assumption}

\noindent Histories not satisfying Assumption \ref{Asmp:Read-After} are considered buggy and will not be considered for consistency measurement.

For any history $\sigma$, since all writes are unique, we can naturally group the operations into clusters:
\begin{definition}[Cluster]
    A cluster $c_i (1\leq i \leq n_w)$ consists of one unique write operation and all its dictated reads.
\end{definition}

\noindent Obviously, the number of clusters is also $n_w$. We assume that the clusters are totally ordered by the start time of the dictating writes, which can be achieved by pre-processing the history in $O(n\log n)$ time.

\subsection{The $i$-atomicity model and the $i$-AV problem}

The basic idea of the $i$-atomicity model is that, when all operations in a history are put into a legal permutation and less than $i$ inversions are incurred, the history is considered $i$-atomic. To give the precise definition, we first define the key notion of an \textit{inversion}. Let $\pi$ denote one permutation of all operations in a history $\sigma$, and let $\pi(i)$ denotes the $i^{th}$ operation in $\pi$. For any permutation $\pi$, we can decide whether any pair of operations form an inversion with reference to $\sigma$ \footnote{The history $\sigma$ is omitted when it is obvious from the context.}:
\begin{equation*}
B_{\pi}(i,j) = \begin{cases}
1 & \text{if $i<j$, but $\pi(j) \rightarrow \pi(i)$ in $\sigma$} \\
0 & \text{otherwise}
\end{cases}
\end{equation*}

\noindent Given $B_{\pi}(i,j)$ for any pair of $i$ and $j$, we can count the maximum number of inversions incurred by one operation in $\sigma$:
\begin{equation*}
\mathcal{I}_{max}(\pi) = \max_{1\leq i \leq n} \left\{ \sum_{1\leq j <i} B_{\pi}(j,i) + \sum_{i <k \leq n} B_{\pi}(i,k) \right\}
\end{equation*}

\noindent After defining what an inversion is and how to calculate the number of inversions of our concern, we can further define $i$-atomicity and $i$-AV:

\begin{definition}[$i$-atomicity]
    For integer $i \geq 0$, the history $\sigma$ is defined to be $i$-atomic, if there exists some permutation $\pi$ of all operations in $\sigma$ such that:
    \begin{itemize}
        \item The permutation $\pi$ is legal, i.e. every read operation reads the value of its latest preceding write in $\pi$.
        \item The history $\sigma$ is not too out-of-order with reference to the legal $\pi$, i.e. $\mathcal{I}_{max}(\pi) \leq i$.
    \end{itemize}
\end{definition}

\noindent The $i$-atomicity model is obviously a generalization of atomicity \cite{Lamport86a, Herlihy90}. When $i=0$, $i$-atomicity transforms to atomicity.

The definition of $i$-atomicity naturally defines the $i$-AV problem:
\begin{definition}[$i$-AV]
\ 
\begin{itemize}
    \item INSTANCE: One history $\sigma$; integer $i \geq 0$.
    
    \item QUESTION: Is $\sigma$ $i$-atomic?
\end{itemize}
\end{definition}

\noindent For the $i$-atomicity model to be useful in practical scenarios, we need to explore efficient algorithms for the $i$-AV problem. One such an attempt is presented in the following Section \ref{Sec:Alg}.

\section{An efficient $i$-AV algorithm for a restricted class of histories} \label{Sec:Alg}

To verify whether history $\sigma$ is $i$-atomic, we need to find a \textit{certificate}, i.e. a permutation with no more than $i$ inversions. Our search for the certificate is basically brute-force enumeration. The search is facilitated by a structure we call the Configuration Graph (CG), inspired by \cite{Golab18}. More importantly, pruning of CG leveraging reasonable assumptions from practical scenarios can reduce the search cost from exponential to polynomial time. We first outline the brute-force enumeration process. Then we define the CG. Finally, we propose the pruning and the cost analysis.

\subsection{Configurations during the search for the certificate}

The clusters are sorted according to the start time of the dictating writes, and are named $c_1, c_2, \cdots, c_{n_w}$ accordingly. To search for the certificate, we scan the (sorted) clusters one by one and the pointer $idx$ ($1\leq idx \leq n_w+1$) slides to the cluster currently being processed. The special index $n_w+1$ is used to indicate the end of the scan. The certificate is constructed by appending all clusters one by one to the prefix $\pi_{pre}$ of the final permutation. The current cluster $c_{idx}$ can be decided, i.e., appended to $\pi_{pre}$, or buffered and decide its position in $\pi_{pre}$ later. The $C_{buf}$ is the (sorted) set of buffered clusters.

According to the state transformation described above, we color the clusters, which is analogous to the coloring used in standard graph traversals:
\begin{itemize}
    \item WHITE: The clusters which have not been processed yet are colored WHITE.
    
    \item BLACK: When the cluster has been appended to $\pi_{pre}$, it is colored BLACK. Its position in the certificate has been decided and will no longer change.
    
    \item GRAY: When the cluster has been processed, but has not been appended to $\pi_{pre}$, it is  colored GRAY. For a GRAY cluster, it is either being processed (i.e., it is $c_{idx}$), or buffered in $C_{buf}$.
\end{itemize}

\noindent Upon appending a new cluster to $\pi_{pre}$, we calculate the maximum number of inversions incurred by any operation in $\pi_{pre}$ so far, and this currently maximum number is recorded in $inv$. Note that during the search, the value of $inv$ never decreases.

\begin{figure}[ht]
    \center
    \includegraphics[width=0.9\columnwidth]{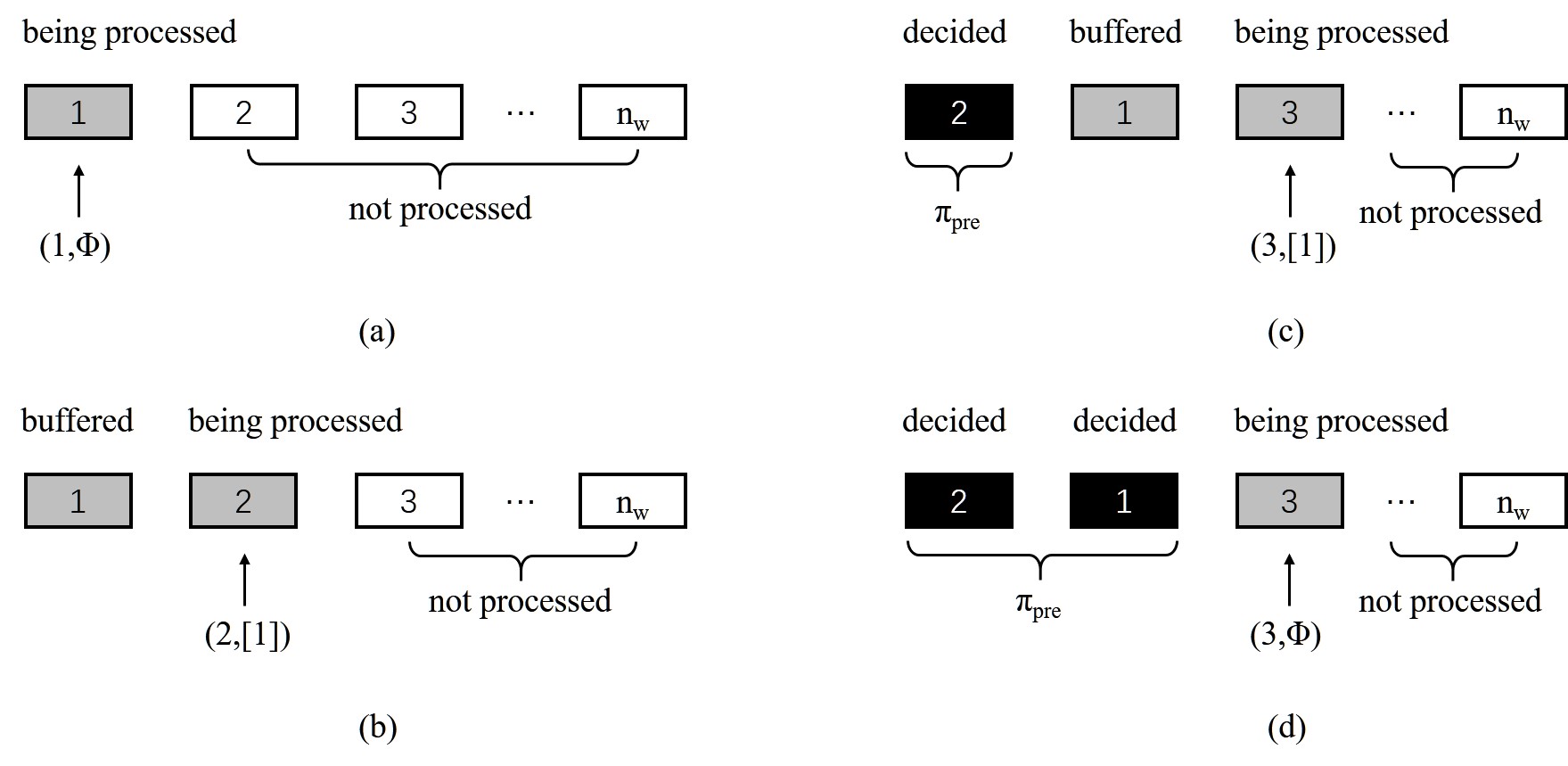}
    \caption{Configurations in the search for the certificate.}
    \label{F:Traversal}
\end{figure}

Summarizing all important information recorded in the search for the certificate, we define the intermediate state of the search a configuration:
\begin{definition}[Configuration]
    The configuration is a quadruple $(idx, C_{buf}, \pi_{pre}, inv)$, describing an intermediate state during the search for the certificate.
%
%
%
\end{definition}


\subsection{The configuration graph}

\begin{figure}[ht]
    \center
    \includegraphics[width=0.6\columnwidth]{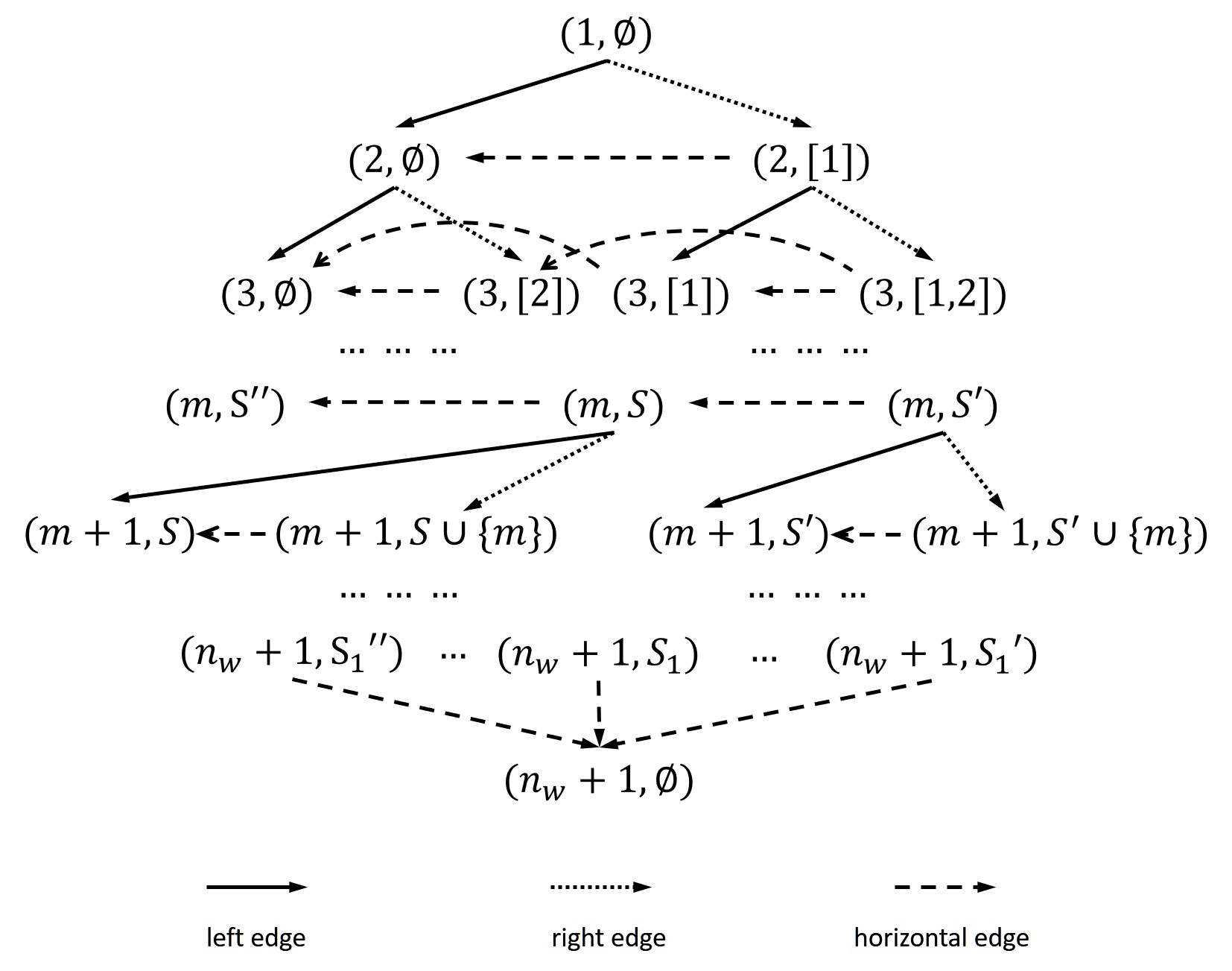}
    \caption{Configuration graph.}
    \label{F:Conf-Graph}
\end{figure}

We organize all the configurations in the Configuration Graph (CG), as shown in Fig. \ref{F:Conf-Graph}. First note that, one node in the CG does not correspond to one configuration. We group all configurations with the same values of $idx$ and $C_{buf}$ (and different values of $\pi_{pre}$ and $inv$) together. A node in CG is defined to represent each group of configurations. 

For the ease of understanding, all nodes with the same $idx$ values (and different $C_{buf}$) are put on the same level in CG. Moreover, two special nodes require further explanations. The search for a certificate always starts from the initial node $v_{ini} = (1, \emptyset)$. The successful search always ends at the final node $v_{final} = (n_w+1, \emptyset)$. Note that in Fig. \ref{F:Conf-Graph}, the special node $v_{final}$ is intentionally drawn in a separate line.

Though one node in CG may correspond to different configurations during the search, the path from $v_{ini}$ to some node $v$ uniquely decides one configuration. To explain this, we first define edges between nodes. Given a node $v = (idx, C_{buf})$, we can process a new cluster by moving the index pointer $idx$ one step rightward (i.e., $idx$ is increased dy 1). The newly covered cluster $c_{idx+1}$ can either be appended to $\pi_{pre}$ or buffered. Meanwhile, we can also stop processing new clusters, but choose some cluster from $C_{buf}$ and append it to $\pi_{pre}$. We denote the three types of processing above by three types of edges stemming from $v$:
\begin{itemize}
    \item Left Edge: The left child is $v_l = (idx+1, C_{buf})$. This means that cluster $c_{idx}$ has been decided and is appended to $\pi_{pre}$. Thus the $idx$ is increased by 1 and the buffer of clusters is not changed.
    
    \item Right Edge: The right child is $v_r = (idx+1, C_{buf} \cup \set{c_{idx}})$. This means that a new cluster is scanned and directly buffered for later processing. The $\pi_{pre}$ is unchanged.

    \item Horizontal Edge: The node may have multiple siblings. The sibling $v_h = (idx, C_{buf } \backslash \set{c})$ has exactly one cluster $c$ taken out from $C_{buf}$ and appended to $\pi_{pre}$.
\end{itemize}

\noindent The way edges are drawn corresponds to their names. Also note that, horizon edges pointing to $v_{final}$ is intentionally drawn downward, due to the special position of $v_{final}$.

First observe that the initial CG node $v_{ini} = (1, \emptyset)$ corresponds to exactly one configuration $(1, \emptyset, \epsilon, 0)$ \footnote{Here $\epsilon$ denotes the empty string. The permutation $\pi_{pre}$ is considered a string of sorted clusters.}. Given explanations of all types of transitions, it is obvious to verify that the path from $v_{ini}$ to $v$ uniquely defines one configuration. For example, consider path $p_1 = (1, \emptyset) \rightarrow (2, \emptyset) \rightarrow (3, \emptyset)$, and path $p_2 = (1, \emptyset) \rightarrow (2,[1]) \rightarrow (3,[1]) \rightarrow (3, \emptyset)$. Let $F(p_1)$ denote the configuration decided by path $p_1$. We can see that though the two paths end at the same CG node, the prefix $F(p_1).\pi_{pre} = \langle 1,2 \rangle$, while $F(p_2).\pi_{pre} = \langle 2,1 \rangle$ \footnote{For the ease of presentation, we use the index value $j$ to denote the cluster $c_j$ when no ambiguity is incurred.}.

\subsection{Traversal and pruning}

It is straightforward to conduct a depth-first search on the CG for the certificate. When we arrive at a node along some path, if one cluster is appended to $\pi_{pre}$, the cost $inv$ is updated. Whenever we find $inv>i$, the node will be decided illegal, and removed from the CG. Obviously, the brute-force traversal of the CG requires exponential time.

The key to improvements on the search cost is the pruning of CG, leveraging  assumptions which are reasonable in real scenarios. Specifically, we assume that:
\begin{definition}[Bounded $i$]
    We assume that the parameter $i$ in $i$-atomicity is part of the input, but its value has a predetermined bound.
\end{definition}

\begin{definition}[Bounded $w$]
    Let $w$ denote the number of concurrent writes in the history. We assume that $w$ has a predetermined bound.
\end{definition}

Though the parameter $i$ in the $i$-atomicity model can be any large integer, we usually are not concerned of too large $i$ values. When there are too many inversions, the quality of the distributed storage service probably has been unacceptable and the quantification is no longer necessary. When we conduct measurement of data consistency, we actually (implicitly) assume that, the data is not too inconsistent. This justifies our assumption on the bound of $i$. Similar phenomenon has also been observed for the $k$-atomicity model. When using Cassandra to provide eventually consistent storage service and measure data consistency for the history, the actual inconsistency measured by $k$ is bounded \cite{Huang18}. Also, many eventually consistent distributed storage services emphasizing high availability and low latency are tuned for read-dominant workloads. The write operations are expensive and are expected to be much less frequent than read. Thus, the concurrency among writes in such workloads is also expected to be bounded.

Given the assumptions above, basic idea of the pruning is that the buffer cannot  grow arbitrarily. If too many clusters are buffered, only a bounded part of the buffered clusters are concurrent (bounded $w$). The other clusters are out of order and can cause too many inversions (exceeding the bound $i$). Thus, CG nodes with oversized buffers are illegal and pruned without processing. This intuition is captured by the following two lemmas:

%
%

%
%
%
%

\begin{lemma}[Bounded buffer sizes] \label{LM:Buffer-Size}

The buffer size of any configuration is bounded by:

\begin{equation*}
    0 \leq |C_{buf}| \leq i+w
\end{equation*}

\end{lemma}

\pf For any configuration with buffer size $i+w$, we prove that the newly covered cluster ``should not" be added to the buffer. Let configuration $F=(idx, C_{buf}, \pi_{pre}, inv)$. Assume that $|C_{buf}|=i+w$. Also assume for contradiction that $F$ transforms to $F'$ by a Right Edge, i.e., the cluster being processed is added to the buffer. Thus, $F'=(idx+1, C_{buf} \cup \set{c_{idx}}, \pi_{pre}, inv)$.

We can show that if $c_{idx}$ is appended to $\pi_{pre}$ before any cluster in $C_{buf}$, at least $i+1$ inversions will be incurred. To see this, first observe that there are at most $w-1$ write operations in clusters in $C_{buf}$ which are concurrent with the write of $c_{idx}$. Thus there are at least $(i+w)-(w-1) = i+1$ writes, which are not concurrent with the write of $c_{idx}$. Since the clusters are sorted by the start time of the dictating writes, and scanned one by one according to this order, the $i+1$ writes which are not concurrent with the write of $c_{idx}$ must precede it. Appending $c_{idx}$ to $\pi_{pre}$ first will incur at least $i+1$ inversions.

Thus, if we want to finally reach $v_{final}$, we must append at least one cluster in $C_{buf}$ to $\pi_{pre}$ before adding $c_{idx}$ to $C_{buf}$. Thus configuration $F'$ will never be reached for the traversal from $v_{ini}$ to $v_{final}$. It can be safely pruned. \pfend

We further find that even the buffer size is legal (no more than $i+w$), the index values of buffered clusters can only appear in a limited range of integers:
\begin{lemma}[Bounded buffer indexes] \label{LM:Buffer-Index}
    
For configuration $F = (idx, C_{buf}, \pi_{pre}, inv)$, the indexes of clusters in $C_{buf}$ can only have values in the given range:    
\begin{equation*}
    C_{buf} \subseteq \set{c_x | idx-2i-2w+1 \leq x < idx}
\end{equation*}
    
\end{lemma}

\pf It is obvious that $x < idx$. What we need to show is that the ``oldest" cluster which can be in the buffer is $c_{idx-2i-2w+1}$. Assume for contradiction that cluster $c_{old} = c_{idx-2i-2w}$ is also in the buffer.
    
\begin{figure}[ht]
    \center
    \includegraphics[width=0.8\columnwidth]{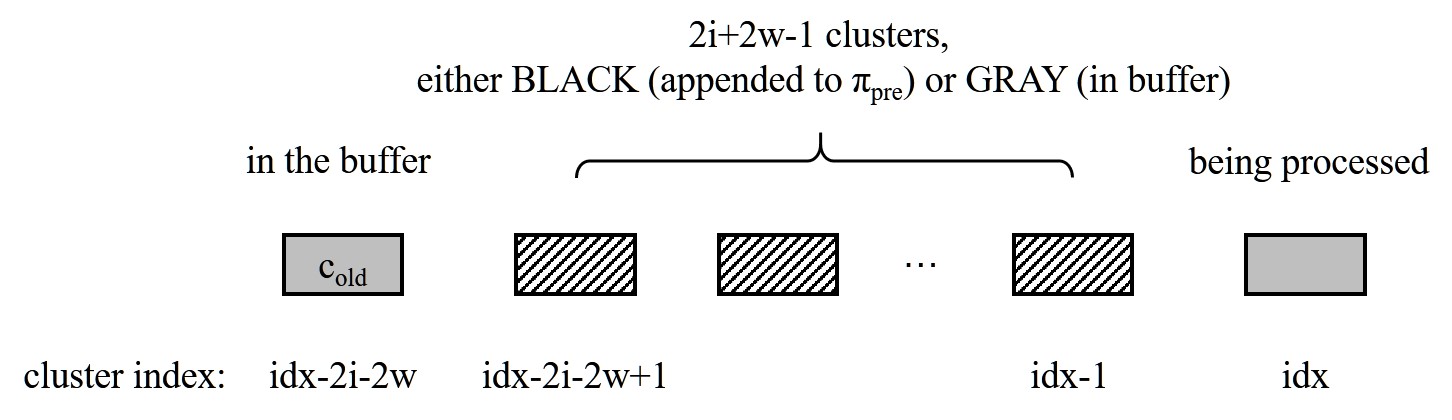}
    \caption{Indexes of clusters in the buffer.}
    \label{F:Buffer-Index}
\end{figure}

Note that the current cluster being processed is $c_{idx}$. Consider all clusters whose index values range from $idx-2i-2w$ to $idx-1$, as shown in Fig. \ref{F:Buffer-Index}. All these clusters have been appended to $\pi_{pre}$ or have been put into the buffer. Denote by $C_1$ all the BLACK clusters which have been appended to $\pi_{pre}$, and by $C_2$ all the GRAY clusters which are in the buffer, excluding $c_{old}$. It is obvious that $|C_1|+|C_2|=2i+2w-1$, and  $C_1 \cap C_2 = \emptyset$.

We first show that $|C_2|\leq i+w-1$. This is because, any buffer size is bounded by $i+w$, according to Lemma \ref{LM:Buffer-Size}. Since $c_{old}$ is also (assumed to be) in the buffer, but it is not included in $C_2$ by definition, the size of $C_2$ is bounded by $i+w-1$.

Based on the bound on $C_2$ size, we have that $|C_1| \geq i+w$. Consider all clusters in $C_1$ and cluster $c_{old}$, we can find more than $i$ inversions incurred. This is because, at most $w-1$ writes of clusters in $C_1$ are concurrent with the write of $c_{old}$. Thus, there are at least $i+1$ remaining clusters in $C_1$ which have been appended to $\pi_{pre}$ before $c_{old}$. Note that $c_{old}$ is still in the buffer. When $c_{old}$ is appended to $\pi_{pre}$ some time later, the clusters in $C_1$ will make $c_{old}$ incur at least $i+1$ inversions, which makes the configuration $F$ illegal. Thus, we have that for any (legal) configuration, the buffer index can only have values in the given range.  \pfend

The bound on buffer sizes and index values help us to bound the size of CG:
\begin{lemma}
    
The CG has no more than $(n_w+1)2^{2i+2w-1}$ nodes, and no more than \\ $(i+w+1)(n_w+1)2^{2i+2w-1}$ edges.
    
\end{lemma}

\pf The CG has $n_w+1$ levels. On each level, the number of nodes are decided by the number of different $C_{buf}$, which is bounded by the number of all subsets of the set of permitted index values in the buffer. According to Lemma \ref{LM:Buffer-Index}, the number of possible index values is bounded by $2i+2w-1$. Thus, there are at most $2^{2i+2w-1}$ different buffers on one level of the CG \footnote{Note that, since the buffer size is no more than $i+w$, this bound can be approximately improved to its half, i.e. $\frac{1}{2}2^{2i+2w-1}$.}. Thus, the number of nodes in CG is bounded by $(n_w+1)2^{2i+2w-1}$.

For each node, there are at most 1 left edge and 1 right edge. The horizontal edge points to configurations with one cluster missing in the buffer. The buffer size is $i+w$. Thus there are at most $i+w$ horizontal edges. The number of edges stemming from one nde is bounded by $i+w+2$. The bound can be slightly improved to $i+w+1$. If the buffer size is $i+w$, this node cannot have the right edge. This corresponds to the case where the buffer is full (as required by Lemma \ref{LM:Buffer-Size}) and cannot accept any more clusters. Thus, the number of edges is bounded by $(i+w+1)(n_w+1)2^{2i+2w-1}$. \pfend

%

\begin{theorem}
    
The cost for solving $i$-AV is $O(nn_w(i+w+1)(n_w+1)2^{2i+2w-1} + n^2) = O(n^3)$.
    
\end{theorem}

\pf The cost for solving $i$-AV is based on the traversal of CG. When processing each node and edge, we need to calculate the maximum number of inversions incurred by any operation so far. To calculate this maximum number, we can pre-process the history $\sigma$ to obtain a table which records the maximum number of inversions that will be incurred when put an operation before and after a given cluster. This pre-processing can be done in $O(n^2)$ time. Given the pre-processing, the update of the maximum number of inversions during the CG traversal can be done in $O(n n_w)$ time.

In summary, the cost for solving $i$-AV is $O(n n_w(i+w+1)(n_w+1)2^{2i+2w-1} + n^2)$. This cost is $O(nn_w^2+n^2) = O(n^3)$, when $i$ and $w$ are bounded. \pfend

%
%
%
%
%
%
%
%
%

\section{Conclusion and future work} \label{Sec:Conclu}

In this work, we study the problem of measuring data consistency for eventually consistent distributed storage systems. An inversion-based metric - $i$-atomicity - and a polynomial verification algorithm for a restricted type of histories are proposed. Our current progresses give rise to a number of important issues to be addressed, concerning the $i$-atomicity model and the $i$-AV problem, as detailed below.

\textit{Variations in defining $i$-atomicity}. Given the basic idea of using inversions to measure data consistency, there are different ways to count the number of inversions. In our model, we use the $\mathcal{I}_{max}(\pi)$ function, which calculates the maximum number of inversion incurred by one operation. The $\mathcal{I}_{max}(\pi)$ function principally means that any single operation cannot cause too much disorder. Another function which is also intuitively reasonable is the $\mathcal{I}_{sum}(\pi)$ function:
\begin{equation*}
    \mathcal{I}_{sum}(\pi) = \sum_{1\leq i < j \leq n} B_{\pi}(i,j)
\end{equation*}

\noindent The $\mathcal{I}_{sum}(\pi)$ function calculates all the inversions incurred in the permutation. It basically means that, the cumulated degree of disorder cannot exceed the user-specified bound. Though both functions are intuitive, our verification algorithm based on the CG traversal cannot be easily applied to the $\mathcal{I}_{sum}(\pi)$ case. The key challenge is that, when we reach the same CG node from a different path, we may lower the number of inversions (calculated by $\mathcal{I}_{sum}$) incurred. This reduction in the number of inversions may requires us to re-check CG nodes which had been determined illegal since the number of inversions goes beyond the user-specified bound. More powerful techniques are required to prune the CG for the $\mathcal{I}_{sum}(\pi)$ function.

\textit{Handling of multiple shared objects}. Existing verification algorithms for the AV and $k$-AV problems are designed for histories with only one shared object. However, both atomicity and $k$-atomicity are local, i.e., for a history with multiple shared objects,  it is atomic/$k$-atomic if and only if for each object accessed, the sub-history is atomic/$k$-atomic \cite{Golab18}. The locality of atomicity and $k$-atomicity makes the assumption of single shared object non-restrictive. Unfortunately, the $i$-atomicity model is not local (which can be easily proved using a counter-example). When the history contains multiple shared objects, our verification algorithm can still conduct the search, but the pruning cannot work. Better techniques are necessary to obtain (pseudo-)polynomial algorithms for the multiple shared object case.
    
\textit{Optimization by segmentation}. The efficiency of our current verification algorithm may be improved by the notion of \textit{forward/backward zones} \cite{Gibbons97}. The basic idea is that using the notion of zones, temporally unrelated parts in the history can be segmented into multiple subhistories. This optimization may not improve the asymptotic growth rate of the verification algorithm, but the improvements in efficiency may be important for practical use of the $i$-atomicity model for consistency measurement.

\clearpage
\bibliographystyle{acm}
\bibliography{inversion}

\clearpage
\appendix


\end{document}